# Probing the X-ray Soft Excess in the Narrow-Line Seyfert-1 (NLS1) Galaxy Tonantzintla S180

B. Korany [1,2]

[1]Department of Astronomy, National Research Institute of Astronomy and Geophysics (NRIAG), 11421 Helwan, Cairo, Egypt
[2]Department of Physics, Faculty of Applied Science, Umm Al-Quran University, Saudi Arabia

**Abstract:** In this work, we present detailed X-ray Soft Excess of the narrow-line Seyfert-1 galaxy Tonantzintla S180 (Ton s180). The study used four XMM-Newton observations for the period from the year 2000 to the year 2016 taken from the XMM-Newton archive. Two different models were used to treat the X-ray Soft Excess for each data separately, the first model is the power law component and the second one is two black body components. We found that, for all observations, the spectra fitted by using a power law component are poor and the output parameters much different from the previous studies, this model unsuccessfully reproduces the continuum shape of the spectrum, including the soft excess, of all observations (Nh = 3.62 ×$10^{20}$ cm$^{-2}$ and photon index for the hard and soft bands are 3.31 and 1.57). The study proved that it can rely upon the two black body components in treating X-ray Soft Excess of the narrow-line Seyfert-1 galaxy (Nh= 1.395 × $10^{20}$ cm$^{-2}$ and The temperature of the two black-body components (KT) are 0.075 and 0.17 keV).

## 1. Introduction

Depending on their optical emission line properties, the Seyfert 1 galaxies are classified into two types: Type 1 is the brood line Seyfert 1 galaxies (BLS1s), in this type Balmer lines (Hα, H$_β$, and Hγ, etc.) are broader than forbidden lines such as [OII], [OIII], their gas density is higher. The full width at half maximum of Balmer lines are more than 2000 km/s and for forbidden lines are less than this value (Muroshima 2005). Type 2 is a narrow line Seyfert 1 galaxies (NLS1s) which show similar line width for Balmer lines and forbidden lines, about hundreds of km/s, and their density is low enough for the forbidden lines can be emitted from.

The best tool to probe matter around the central black hole in the Seyfert galaxies is the X-ray spectral components and features. The AGN spectra contain several components, such as power law component, cold absorber, partial absorber, warm absorber, a soft X-ray excess, iron K lines, and hard X-ray hump. The photon energy distribution near the central black holes is power law distribution which leads to the power law components. The iron K lines and hard X-ray hump are interpreted as fluorescent X-rays by irradiation of power law continuum to the accretion disk. The soft excess is a major component of the X-ray spectra of many AGN. Various sophisticated models have been proposed to explain the soft X-ray excess component. Generally, the X-ray soft excess is modeled empirically by a single blackbody component with a temperature of 0.1-0.2 keV. Adding on the blackbody model there are other different models such as multiple black bodies, multicolor dick black body, double power-law model, blurred reflection from partially ionized material, smeared absorption, and thermal comptonization in the optically thick medium in the X-ray (Korany and Nouh. 2019). The relativistic reflection model is the leading method of estimating spin in AGN (Fabian et. al 1989). In this technique, the X-rays produced in the X-ray corona illuminate the inner accretion disk and produce the relativistic distortion of narrow emission lines.

Ton S180 is one of the X-ray brightest ultra-soft Seyfert galaxies with narrow lines (NLS1) and low Galactic column density, approximately $N_H = 1.5 \times 10^{20} cm^{-2}$ (Dickey & Lockman 1990) and have redshift z=0.06198 (Wisotzki et al. 1995). The source is at the extreme end of the Seyfert range of line widths with FWHM Hα and Hβ ~ 900 Km s$^{-1}$, making it a good choice for isolating the fundamental parameter that determines the classification of a Seyfert galaxy (Turner et al 2002). This object was observed in X-ray by many X-ray satellites, as ROSAT in 1995, Beppo-SAX in 1998, Chandra in 2001and 2004, ASCA in 2002, RXTE in 2004, and XMM-Newton 2000, 2002, 2015, and 2016. Fink et al. (1997) analyzed ROSAT observation and identified this object as a bright X-ray source detected in all sky survey. Along observation was performed with ASCA in 1999 during a simultaneous campaign with pointing with RXTE. The Ton S180 X-ray soft excess is unusual (see Fig 1), and there is no clear break indicating the end of soft excess and start power-law continuum (Parker et al 2017). Turner et al. (1998) fitted spectra from ASCA observation in the energy range 0.6-10.0 keV with power law component of photon index 2.5 keV plus a broad Gaussian emission line cantered at 0.82 keV. Also; Comatrie et al. (1998) fitted spectra for Beppo-Sax 1996 observations in the energy range 0.1-10.0 keV with double power law component for photon indices 2.3 and 2.7 keV. The same model (double power law) was used by Vaughan et al in 2002 with XMM-Newton observation observed in 2000. The photon indices used by Vaughan et al was 3.1 and 1.5 keV which is slightly different than those used by Comatrie et. al in 1998.

We report in this paper on X-ray Soft Excess in the Narrow-Line Seyfert-1 (NLS1) Galaxy Ton S180, by studying four XMM-Newton observations over sixteen years. We organized the paper as follows: the observations and data reduction appears in section 2, section 3 is devoted to the spectral analysis and modeling, and the conclusion is outlined in section 4.

**2- Observation and Data Reduction**

The observational data files (ODFs are retrieved from the XMM–Newton public archive. The object Ton S180 was observed four times:

1- In 2000 December 14 for a duration of approximately 30 KS. MOS2 and pn cameras were operated in a small window mode to reduce the overexposure of the CCD, occurring when two or more independent photons arrived at nearby pixels within one read-out cycle (pile-up). MOS1 was in timing mode.

2- In,2002 Jun 30 for a duration approximately 18 KS. The MOS1 in timing mode, and MOS2 and pn in small window mode.

3- In 2015 Juley 3 for a duration of approximately 14 KS. In this observation all three European Photon Imaging Cameras (EPIC) in small window mode.

4- 2016 Juley 13 with all three EPIC cameras in small window mode. All the observations were observed in medium filter except MOS2 and pn cameras of 2015 observation in thin filter1.

Table 1 the observations log of all the data. The raw data for all observations were processed with the EPIC pipeline chains of the Science Analysis System (SAS) software version 13.0.0., to produce calibrated event list files and screened to remove unwanted pixels. The event files of the individual observations have been cleaned from some bad time intervals characterized by high background events (so-called soft-proton flares). These bad time intervals are rejected by creating light-curves for the observations (PN, MOS1, and MOS2), which are best visible above 10 keV. The spectra of the object were extracted using circular extraction regions cantered on the object, with a radius of 35 arc-sec for both the MOS and PN cameras. We used the same circular size from the source-free region for the background spectra (Korany& Nouh 2019). All the spectra were subsequently analyzed using the **Xspec** software (v12.9.0). Due to its larger effective area and spectral range, we used the pn data only in the subsequent spectral analysis.

Table 1: XMM-Newton Observations log (MOS1,MOS2 & PN ) for TON S 180.

| Obs. year | nstr. | Mode | Filter | Expo. Start Time | Expo. End Time |
|---|---|---|---|---|---|
| 2000 | MOS1 | Timing mode | MEDIUM FILTER | 2000-12-14@11:31:43 | 2000-12-14@19:39:37 |
|  | MOS2 | Small-window | MEDIUM FILTER | 2000-12-14@11:21:16 | 2000-12-14@19:39:10 |
|  | PN | Small-window | MEDIUM FILTER | 2000-12-14@11:35:18 | 2000-12-14@19:46:21 |
| 2002 | MOS1 | Timing Mode | MEDIUM FILTER | 2002-06-30@03:01:39 | 2002-06-30@08:02:49 |
|  | MOS2 | Small-window | MEDIUM FILTER | 2002-06-30@03:01:43 | 2002-06-30@08:07:13 |
|  | PN | Small-window | MEDIUM FILTER | 2002-06-30@03:06:35 | 2002-06-30@08:07:28 |
| 2015 | MOS1 | Small-window | MEDIUM FILTER | 2015-07-03@22:26:17 | 2015-07-05@13:22:38 |
|  | MOS2 | Small-window | THIN FILTER 1 | 2015-07-03@22:27:01 | 2015-07-05@13:22:43 |
|  | PN | Small-window | THIN FILTER 1 | 2015-07-03@22:32:28 | 2015-07-05@13:22:4 |
| 2016 | MOS1 | Small-window | MEDIUM FILTER | 2016-06-13@05:36:14 | 2016-06-13@14:10:55 |
|  | MOS2 | Small-window | MEDIUM FILTER | 2016-06-13@05:36:58 | 2016-06-13@14:11:00 |
|  | PN | Small-window | MEDIUM FILTER | 2016-06-13@05:42:25 | 2016-06-13@14:11:15 |

3- **Spectral Analysis and Results**

The present treatment was carried out by making a spectral analysis for every observation in the four data observations individually. We started the spectral analysis by fitting the hard band (2.5-10 keV) spectra to exclude the soft x-ray excess and to determine the power law photon index and the value of the iron emission line peak. A simple power-law model modified by absorption from the cold gas in our Galaxy is fixed at Nh=1.36 X $10^{20}$ cm$^{-2}$ (which calculated using the Nh command in ftools software). To check the absorption due to the cold gas in the source galaxy, a redshifted cold absorption component was added to the model with a fixed redshift at 0.0619. This component didn't add any improvement to the fitting, and it is clear that from the fitting there is no inner absorption. A simple Gaussian line component for the narrow iron emission due to e.g. reflection from a cold torus used($A(\lambda) = k \, (1/\sigma^2 \sqrt{2\pi}) \, e^{(-((\lambda-\lambda_l)^2/2\sigma^2))}$)

The results of the spectral analysis are summarized in Table 2. In this table, M1 is the model of power law and Gaussian emission line in the hard band (2.5-10.0 keV); M2 is two power law components and Gaussian emission line in broad-band spectrum (0.35-10.0 keV); M3 is two black body components, power law component and Gaussian emission line in the broad-band spectrum (0.35-10.0 keV). The cold absorption from our galaxy component added to all the models.

The power law photon index ($\Gamma$) for the spectra of the four observational data is between $2.25 \pm 3.7 \times 10^{-02}$ and $2.06 \pm 5.7 \times 10^{-2}$. The calculated mean value of photon index for the four spectra the value is $2.14 \pm 4.23 \times 10^{-02}$. The emission line is between $6.6 \pm 4.6 \times 10^{-02}$ and $6.35 \pm 4.96 \times 10^{-02}$ keV and the mean value is 6.48 keV. The reduced $\chi^2$ for the fit of the data of the year 2000 as an example is 0.9067 for 91 degrees of freedom, with $7.3 \times 10^{-01}$ Null hypothesis probability (Fig. 2).

The fitted model (power law and Gaussian line with absorbed column density) of the hard X-ray band (2.5 - 10.0 keV) is extrapolated to the broadband spectral data (0.35 - 10 keV), a huge soft X-ray excess was found (Fig 3).

At the beginning of the soft X-ray excess treatment, the soft band of the spectra is modeled as a power law. The spectra are fitted by a power law model for the soft band (0.35-2.0 keV) with a power law model for the hard band and a Gaussian component for the narrow iron emission line. All the model component parameters lift free during the fit. A component for the partial covering absorption by partially ionized material may be formed by photoionised evaporation from the inner edge of the torus (warm absorber) checked. This absorption is checked by adding to the model a redshifted absorption edge component (the redshift is 0.0619) due to the absorption from the source galaxy [ $M(E) = e^{[-D(E(1+z))/E_c]^3}$ ] (zedge in Xspec), and the absorption edge component due to the gas near our galaxy[$M(E) = e^{[-D(E/E_c)^{-3}]}$] (edge in Xspec) added to the model. No significant improvement occurred for the fit by adding the warm absorber components. The two photon index of the fit for the hard and soft band is from 3.19 to 3.42   and from 1.48 to 1.65 (Table 2) and the mean value is   3.31 and 1.57. The values of the output parameters of the cold absorption from our galaxy for the four spectra are 4.14×10$^{20}$ ± 3.84.1×10$^{-03}$, 3.18×10$^{20}$ ± 4.20×10$^{-03}$, 3.16×10$^{20}$ ± 2.08 ×10$^{-03,}$ and 3.36×10$^{20}$ ± 7.11×10$^{-03}$ cm$^{-2}$ respectively. The mean value of the cold absorption parameter is 3.62 ×10$^{20}$ cm$^{-2}$ which is nearly three times the value calculated from Nh command of ftools application (1.36×10$^{20}$) and the value used by J. Crummy et. al. (2006) (1.55 ×10$^{20}$). We fitted the iron emission line and obtained a poor fit for all observations, for example, the reduced $\chi^2$ of the spectrum of the year 2015 is 1.434 for 150 degrees of freedom, and the null hypothesis probability is 1.688 ×10$^{-05}$ (Figure 4). We went through another scheme by using a model consists of two black-body components used for the soft excess instead of the power law model.

By using the single black-body model for the soft excess for broad energy band (0.35-10.0 keV), with a power law model for the hard band and a Gaussian component for the narrow iron emission line.  The black-body temperature is about 118 ± 16 eV and the power law photon index ($\Gamma$) is 2.498 ±1.365 × 10$^{-02}$. This model provided a poor fit to the data, which give a very high reduced $\chi2$ value.  A Significant improvement occurred for the fit by adding another black-body component to the model for the soft excess (the F test value between the two models is 30.26 for observation of the year 2016).  A Gaussian component for the narrow iron emission line, with a free parameter for the absorption from the cold gas in our Galaxy added to the model.  The $\chi^2$ value for the new fit is improved (for example the value is 144.31 for 142 degrees of freedom for the spectrum of the year 2000) (Fig .5). The Nh absorption of the final model is 1.55 × 10$^{20}$ ± 6.15 × 10$^{-03}$, 1.86 × 10$^{20}$ ± 6.98 × 10$^{-03}$, 1.0 × 10$^{20}$ ± 3.37 × 10$^{-03}$ and 1.35 × 10$^{20}$ ± 7.20 × 10$^{-03}$ cm$^{-2}$ respectively. The mean value for the Nh is 1.395 × 10$^{20}$ this value greatly matches with the output value from the Nh Ftools command (1.36 × 10$^{20}$) and the value used by J. Crummy et. al (1.55 ×10$^{20}$). The peak of the iron emission line has occurred at 6.604 ± 4.5 × 10$^{-02}$, 6.45± 0.06, 6.35± 4.96× 10$^{-02}$ and 6.52± 0.10 keV respectively. The temperatures of the first black-body component (KT) are 0.078 ± 2.89 × 10$^{-3}$, 0.073 ± 3.51×10$^{-3}$, 0.078± 1.99 ×10$^{-3}$ and 0.072 ±3.70 ×10$^{-3}$ keV, with mean value 0.075 keV, and the temperature of the second black-body component(KT) is 0.17 keV for all spectra (Table 2 summarized the output of the all parameters).

Table 2: The summary of the output parameters for the different models

| Obs. ye | model | Nh × 10$^{20}$ | Γ | Γ | KT1 keV | KT2 | ga | σ | Reduced χ$^2$/dof |
|---|---|---|---|---|---|---|---|---|---|
| 2000 | M1 | Freeze | 2.25± 3.7×10$^{-2}$ | | | | 6.6 ± 4.6×10$^{-02}$ | 0.1 ± 5.9×10$^{-2}$ | 0.907/91 |
| 2000 | M2 | 4.14± 3.84×10$^{-3}$ | 3.42 ± 5.8×10$^{-2}$ | 1.6 ± 8.15×10$^{-2}$ | | | 6.6 ± 4.77×10$^{-02}$ | 6.39×10$^{-2}$ ± 9.96×10$^{-3}$ | 1.434/150 |
| 2000 | M3 | 1.55 ± 6.151× 10$^{-3}$ | 2.31 ± 2.46 × 10$^{-2}$ | | 0.078± 2.89 × 10$^{-3}$ | 0.17 ± 6.13 × 10$^{-3}$ | 6.604 ± 4.5 × 10$^{-2}$ | 0.2 ± 7.17×10$^{-3}$ | 0.94/142 |
| 2002 | M1 | Freeze | 2.06 ± 5.7×10$^{-2}$ | | | | 6.42 ± 0.17 | 0.33± 0.02 | 0.7 / 79 |
| 2002 | M2 | 3.18± 4.194×10$^{-2}$ | 3.26 ± 7.7×10$^{-2}$ | 1.53 ± 0.12 | | | 6.37 ± 0.16 | 0.32± 0.02 | 1.0538 / 132 |
| 2002 | M3 | 1.86 ± 6.98×10$^{-3}$ | 2.22 ± 3.23×10$^{-2}$ | | 0.073 ± 3.51×10$^{-3}$ | 0.17± 7.51 × 10$^{-3}$ | 6.45± 0.06 | 0.10 ± 0.01 | 1.1 / 131 |
| 2015 | M1 | Freez | 2.15 ± 2.35×10$^{-2}$ | | | | 6.35± 0.05 | 0.30 ± 0.06 | 1.0591 / 111 |
| 2015 | M2 | 3.16± 2.08×10$^{-3}$ | 3.19 ± 3.45×10$^{-2}$ | 1.48 ±5.52 ×10$^{-2}$ | | | 6.311± 5.66×10$^{-2}$ | 0.27± 6.24×10$^{-2}$ | 2.013/163 |
| 2015 | M3 | 1.0± 3.37×10$^{-3}$ | 2.26± 1.57×10$^{-2}$ | | 0.078± 1.99 ×10$^{-3}$ | 0.17± 4.36 ×10$^{-3}$ | 6.35± 4.96×10$^{-2}$ | 0.30± 5.43×10$^{-2}$ | 1.301/161 |
| 2016 | M1 | Freeze | 2.1 ± 5.3×10$^{-2}$ | | | | 6.53± 0.11 | 0.21± 0.12 | 1.131/84 |
| 2016 | M2 | 3.36± 7.11×10$^{-3}$ | 3.36± 0.11 | 1.65 ± 0.11 | | | 6.52± 0.10 | 0.19± 0.11 | 1.19/135 |
| 2016 | M3 | 1.35± 7.20×10$^{-3}$ | 2.24 ± 3.62×10$^{-2}$ | | 0.072 ± 3.70 ×10$^{-3}$ | 0.17 ± 8.34 ×10$^{-3}$ | 6.52± 0.10 | 0.20± 0.11 | 1.15/133 |

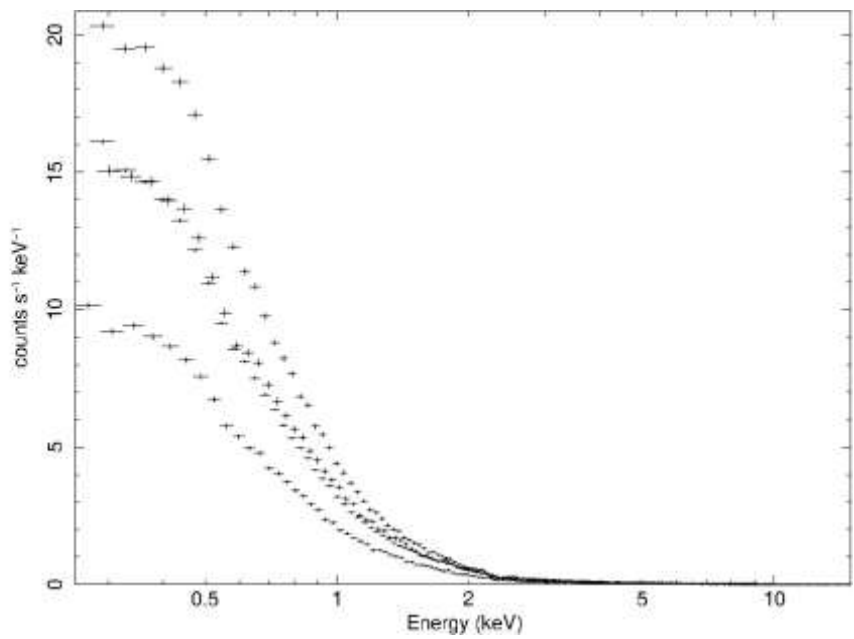

**Fig 1:** The soft excesses of all data of the source

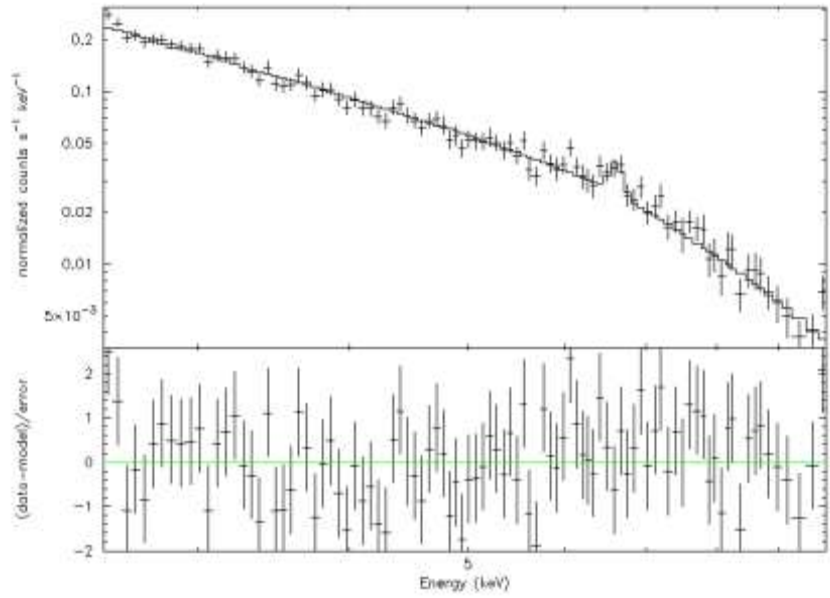

Fig. 2: Spectrum of power law and Gaussian line with cold absorption from our galaxy in the hard energy band (2.5-20.0 keV)

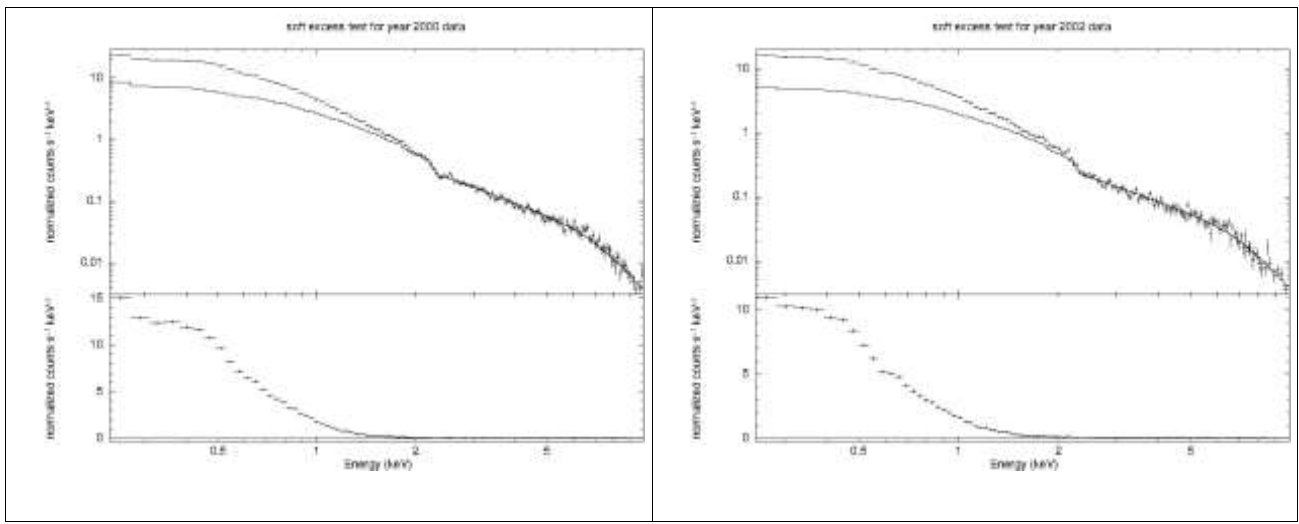

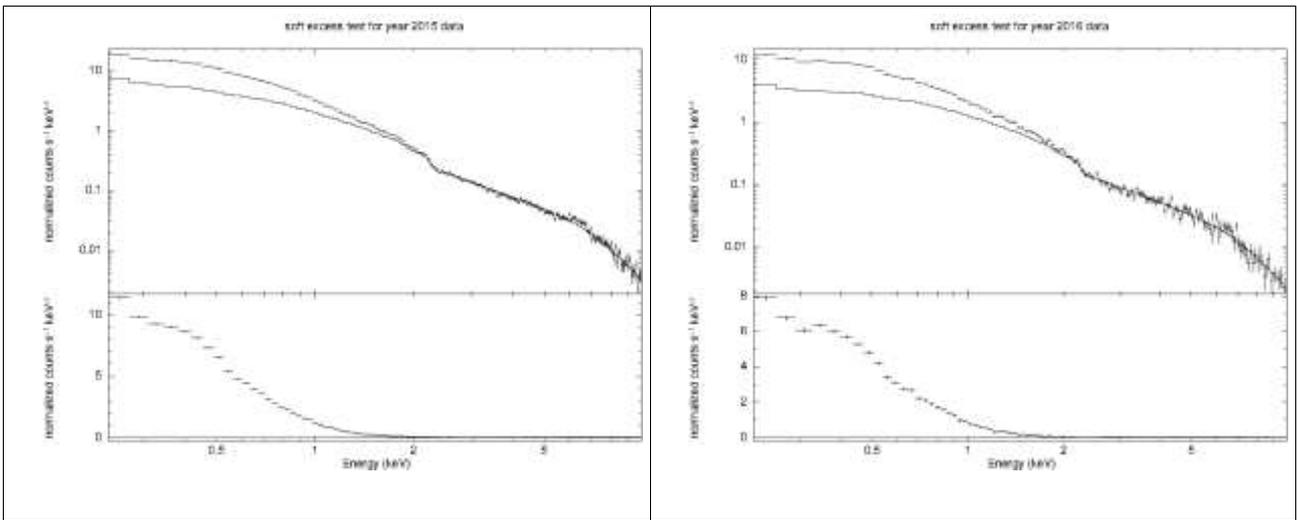

Fig. 3: Extrapolated spectra for broad band (0.35-10.0 keV) with power law and Gaussian line with cold absorption from our galaxy in the hard energy band (2.5-20.0 keV) and without fitting to the soft band.

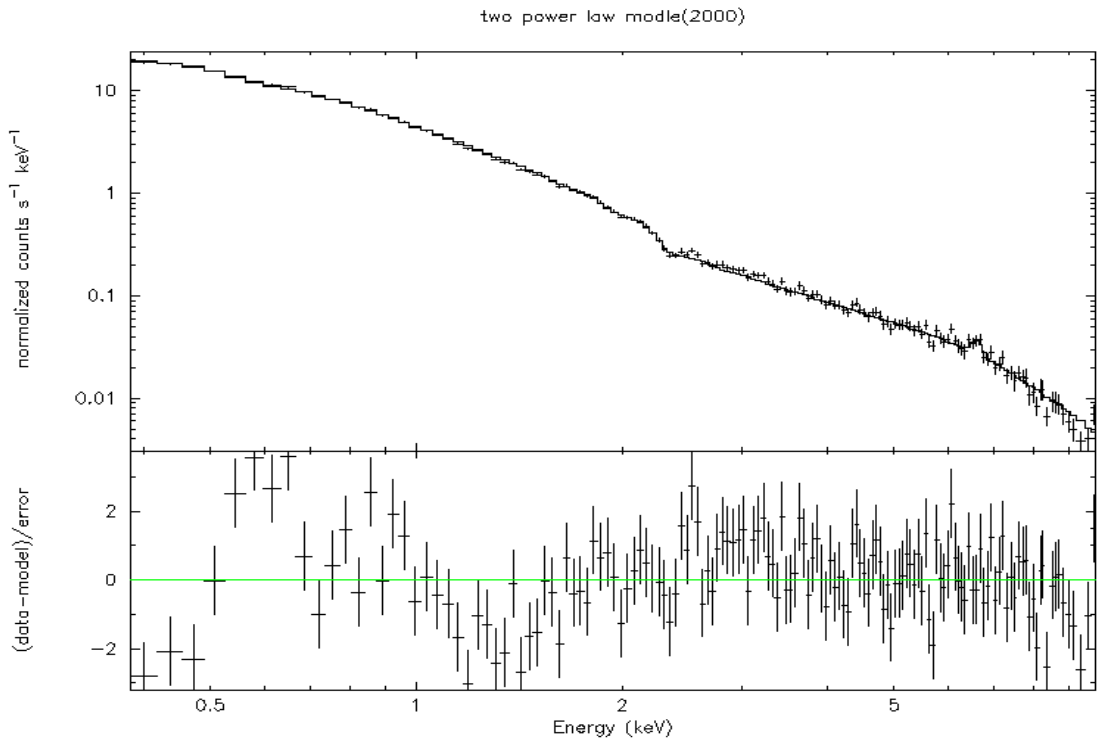

Fig. 4: The two power law component model with cold absorption from our galaxy.

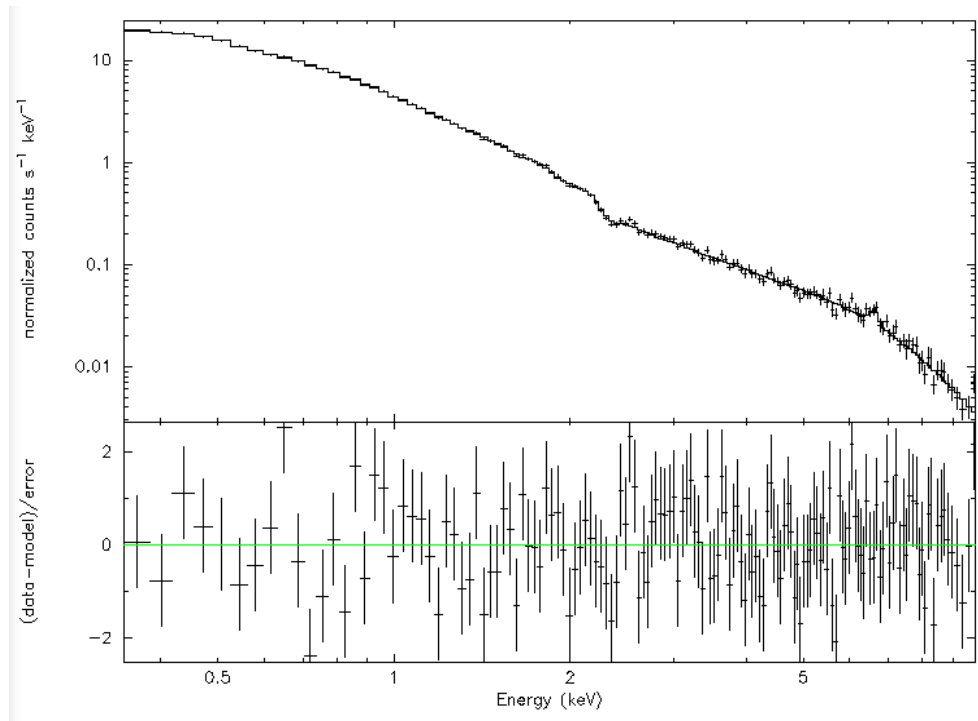

Fig. 5: A cold absorption from our galaxy component with the two black body component models for the soft excess and power law with Gaussian line for hard band.

## 4- CONCLUSIONS

Construction of the SED for the bright NLS1 galaxy Ton S180 shows that most of the energy is emitted in the 10–100 eV regime, indicating that the primary source of emission dominates that band (T. J. Turner rt. al. 2002) and because the soft X-ray excess of the AGN, origin, and nature is still an open issue, we performed a detailed X-ray Soft Excess for a bright narrow-line Seyfert-1, Ton S 180 galaxy. We used observations for a long paired of time, form the year 2000 to the year 2016. The XMM-Newton observed this object in this interval of time four times. In the present spectral analysis, we checked two different models for the X-ray soft excess. The first one is the power law model as advised by several authors (eg. Vaughan et al. (2002)), but this model provided a poor fit to all observational data. The output parameters different from others' results, as the parameter of the absorption from the cold gas in our Galaxy ($3.62 \times 10^{20}$ cm$^{-2}$) and the photon index for the power low of photon energy distribution for the soft and hard bands (3.31 and 1.57). Because the soft X-ray excess is the most probably due to a higher temperature of the accretion disk as often seen in NLS1s(Th. Bolle, et. al. 2003), we used the model of the two black body components. This model provided an excellent fit to all the spectra and provided reliably output parameters. The cold absorption due to the hydrogen column density of our galaxy in the line of sight is $1.395 \times 10^{20}$ cm$^{-2}$. The temperature of the two black-body components (KT) is 0.075 and 0.17 keV, and as revealed from several AGN, if the soft excess is fitted with a black body the temperature of the black body component is 0.1–0.2 (Gierliński, M. and Done, C. 2004). The peak of the iron emission line of the spectral data is at 6.48 keV. According to the result of the spectral analyses and the output parameters of the fitting, we can conclude that the two black-body component models still a reliable model for X-ray soft excess of the AGN.

# REFERENCES

1- Boller, Th.; Tanaka, Y.; Fabian, A.; Brandt, W. N.; Gallo, L.; Anabuki, N.; Haba, Y.; Vaughan, S. Monthly Notice of the Royal Astronomical Society, Volume 343, Issue 4, pp. L89-L93. 08/2003.

2- Comastri, A.; Vignali, C.; et al. Abstracts of the 19th Texas Symposium on Relativistic Astrophysics and Cosmology. 12/1998.

3- Crummy, J.; Fabian, A. C.; Gallo, L.; Ross, R. R., Monthly Notices of the Royal Astronomical Society, Volume 365, Issue 4, pp. 1067-1081. 02/2006.

4- Dickey, John M.; Lockman, Felix J. Annual Rev. Astron. Astrophys., Vol. 28, p. 215-261 (1990).

5- Fabian, A. C.; Canizares, C. R.; Barcons, X Monthly Notices of the Royal Astronomical Society (ISSN 0035-8711), vol. 239, p. 15P-18P.(Aug. 1, 1989)

6- Fink, Uwe; Fevig, Ron A.; Tegler, Stephen C.; Romanishin, William, Planetary and Space Science, Volume 45, Issue 11, p. 1383-1387. 11/1997.

7- Gierliński, Marek; Done, Chris, Monthly Notices of the Royal Astronomical Society, Volume 349, Issue 1, pp. L7-L11, 03/2004.

8- Korany, B. and Nouh, M. I., Astrophysics, Vol. 62, No. 3, September, 2019.

9- Murashima, Mio; Kubota, Aya; Makishima, Kazuo; Kokubun, Motohide; Hong, Soojing; Negoro, Hitoshi, Publications of the Astronomical Society of Japan, Vol.57, No.2, pp. 279-285 04/2005.

10- Parker, M. L.; Miller, J. M.; Fabian, A. C.,Monthly Notices of the Royal Astronomical Society, Volume 474, Issue 2, p.1538-1544, 02/2018

11- Turner, T. J., Romano, P. B., Kraemer, I. M. George, T. Yaqoob, D. M. Crenshaw, J. Storm, D. Alloin, D. Lazzaro, L. Da Silva, J. D. Pritchard, G. Kriss, W. Zheng, S. Mathur, J. Wang, P. Dobbie, and N. R. Collins The Astrophysical Journal, 568:120–132, 2002 March 20

12- Turner, T. J.; George, I. M.; Nandra, K., The Astrophysical Journal, Volume 508, Issue 2, pp. 648-656. 12/1998

13- Wisotzki, L.; Dreizler, S.; Engels, D.; Fink, H. -H.; Heber, U., Astronomy and Astrophysics, v.297, p. L55, 05/1995.